\documentclass[]{IEEEtran}

\IEEEoverridecommandlockouts
\input epsf
\usepackage{graphicx}
\usepackage{url}
\usepackage{xcolor}
\usepackage{upgreek}

\usepackage[colorlinks=true,linkcolor=blue,urlcolor=blue,citecolor=blue]{hyperref}
\usepackage[colorinlistoftodos]{todonotes}

\begin{document}

\IEEEpubid{\makebox[\columnwidth]{979-8-3315-4683-0/26/\$31.00~\copyright2026 IEEE \hfill} \hspace{\columnsep}\makebox[\columnwidth]{ }}

\title{\LARGE Status of the Tsinghua Tabletop Kibble Balance}

\author{Shisong Li$^\dagger$, Weibo Liu, Nanjia Li, Elsayed E.E. Qupasie, Yongchao Ma,\\ Kang Ma, Wei Zhao, Lisha Peng, Songling Huang, Xinjie Yu\\
Department of Electrical Engineering, Tsinghua University, Beijing 100084, China\\
$^\dagger$Email: shisongli@tsinghua.edu.cn} 

\maketitle

\IEEEpubidadjcol

\begin{abstract}
This paper reports on the status of the Tsinghua tabletop Kibble balance experiment, aiming to deliver a mass calibration instrument for kilogram realizations in accordance with the new International System of Units (SI). Major progress since 2024 in different aspects, i.e., electrical, magnetic, mechanical, and optical, is presented. The primary weighing and velocity measurement results are discussed.
\end{abstract}


\pagenumbering{gobble}

\section{Introduction}
The Kibble balance~\cite{Kibble1976} has emerged as one of the most critical instruments for realizing mass standards in the new International System of Units (SI). Currently, high-precision Kibble balances are typically large and complex to operate~\cite{NISTInvitedArticle2016, NRC2017}. In recent years, however, tabletop Kibble balance experiments have garnered increasing attention from both national metrology institutes (NMIs) as next-generation quantum mass standards~\cite{NIST_QEMMS, PTB2025, NPLCPEM2024} and industry as direct standards~\cite{NISTTabletop2024}. Despite this progress, scaling down the experiment may introduce challenges in measurement uncertainty~\cite{Li_MagneticUncertainties}, making the development of compact, high-accuracy Kibble balances a continuing challenge. The Tsinghua tabletop Kibble balance aims to achieve a compact, robust, cost-effective, and user-friendly Kibble balance, with a relative measurement uncertainty of a few parts in $10^8$ at 1\,kg, making it suitable for NMI-level mass calibration~\cite{THUdesign2022}. This report provides an overview of recent progress in four key technical areas: magnetic system design, mechanical structure, optical sensing, and electrical systems.

\section{Magnetic}
The Tsinghua tabletop Kibble balance employs a BIPM-type magnetic circuit with two novel modifications~\cite{li2024magnet}: the inner yoke compensation considerably increases the range of field uniformity along the vertical axis. An optimal splitting design significantly simplifies the open/close operation and enables in-situ operation. The magnet circuit has been finalized, and these concepts have been applied to smaller and larger magnets. A two-step yoke compensation method is proposed to further improve the symmetry of the magnetic profile~\cite{li2025approachrestoringmagneticfield}. Since a magnet-moving mechanism is desired for the velocity measurement, the effect on the external magnetic field must be experimentally evaluated. A recent work in our group has, for the first time, experimentally evaluated this effect, and it has been found that the self-shielding of the present motor configuration can help suppress the external flux error to the $10^{-9}$ order~\cite{TsinghuaMag2025}. The change in magnetic field during the weighing and velocity measurement phases is significant in tabletop Kibble balance experiments~\cite{Li_MagneticUncertainties}, especially those operated with a two-mode, two-phase measurement scheme. One-mode measurement has been shown to be an efficient way to address this effect. A new current ramping scheme that maintains the coil's ohmic heating constant across two measurement phases is proposed and verified through both theoretical analysis and experimental tests~\cite{liu2025}. 

\begin{figure}
    \centering
    \includegraphics[width=0.76\linewidth]{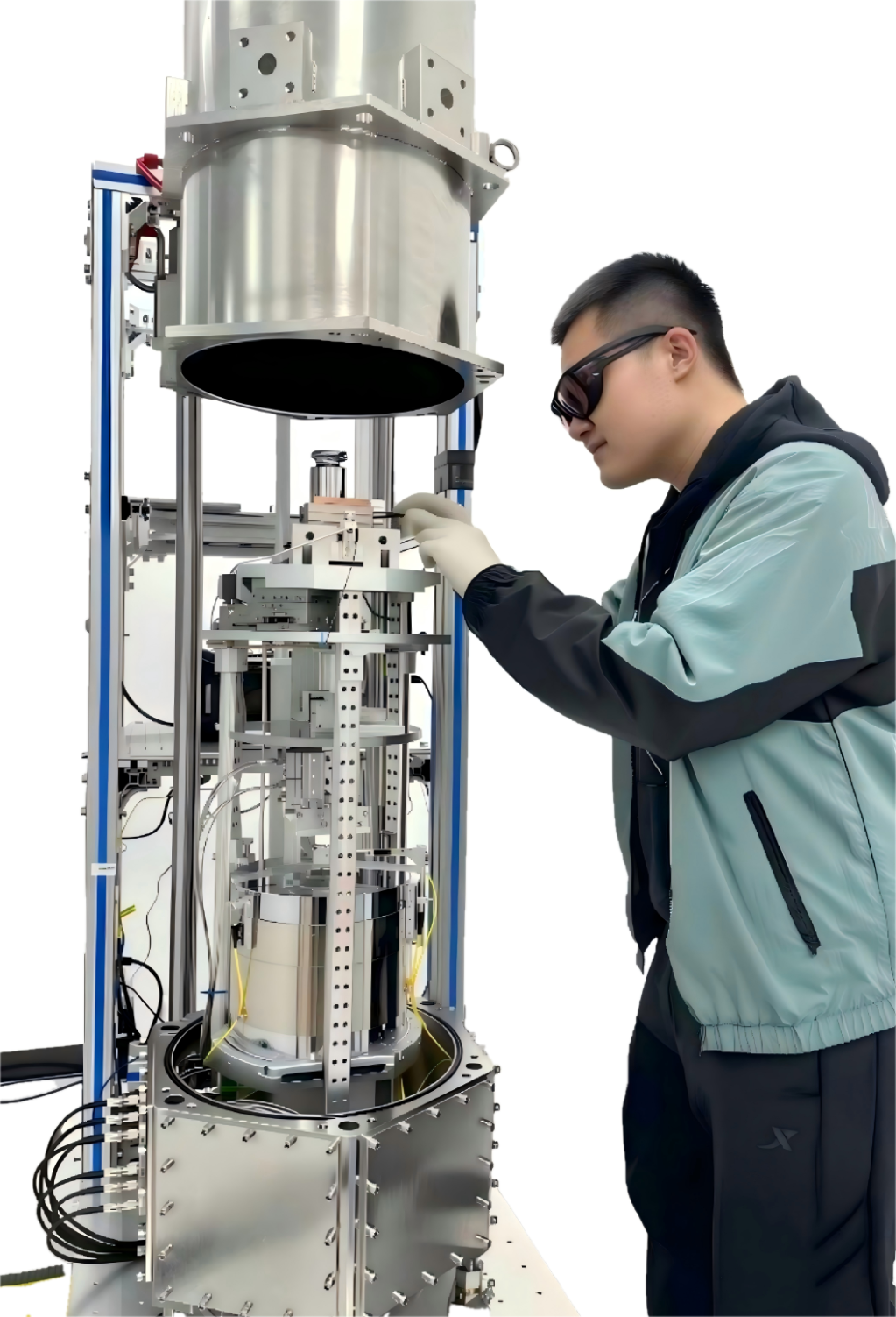}
    \caption{The experimental apparatus of the Tsinghua tabletop Kibble balance. }
    \label{fig:KB}
\end{figure}

\section{Mechanical}
The weighing unit of the Tsinghua system employs a capacitor sensor to detect displacement~\cite{THU2025}. The capacitor sensor offers higher measurement resolution than conventional optical sensors, thereby increasing weighing resolution. This enables a stiffer flexure design with higher load capacity and greater robustness. At present, the weighing phase is operational with a novel PID strategy: a varying but smooth integration term is introduced, which ensures rapid current ramping during the mass-on and mass-off transitions and maintains good stability at static conditions. A preliminary test of the weighing with the 1\,kg standard mass in air shows a peak-to-peak noise around $2\times10^{-6}$ with each mass-on or mass-off measurement lasting 60\,s. A repeatability of less than $1\times10^{-7}$ is achieved within 2 hours. The mechanical structure is now finalizing, as shown in Fig.~\ref{fig:KB}. At present, alignment is underway, and the next step will be measurements in vacuum.

\section{Electrical}
\begin{figure}[h]
    \centering
    \includegraphics[width=\linewidth]{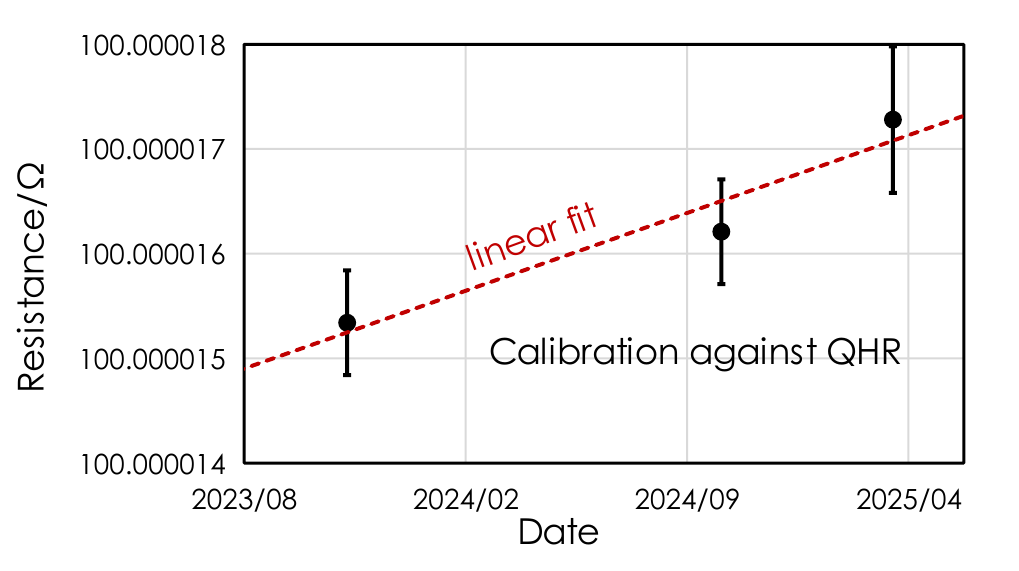}
    \caption{Calibration result of the $100\,\Omega$ standard resistor.}
    \label{fig:R}
\end{figure}

A two-stage digital feedback current source has been built, and short-term stability can reach nA/A with about 30\,mins~\cite{THU2024currentsource}. This current source operates in an open-loop manner during weighing, providing sufficient resolution for coil position control. During the velocity measurement, the current source operates in constant-current mode, making the current effect easy to correct. A cryo-cooled programmable Josephson voltage standard (PJVS) has been developed, providing a precise voltage reference up to 10\,V. A 100 $\Omega$ standard resistor (Alpha-HRU-100), operated in a temperature-controlled oven ($\Delta T<5$\,mK), is employed for the current sampling. As shown in Fig.~\ref{fig:R}, three calibrations of this standard resistor against the QHR system at the National Institute of Metrology (NIM, China) have been carried out. The calibration shows a smooth change in resistance, and the residual is below $1\times10^{-8}$. 

\section{Optical}
Major progress on the interferometer has been made. A fiber-coupled, spatially separated-beam heterodyne interferometer is built inside the vacuum chamber. Two AOM-shifted optical frequencies (150\,MHz and 147\,MHz) generate two Doppler-shifted beat signals, which are detected with a time-interval analyzer (TIA) to determine the relative vertical velocity of the coil with respect to the magnet while suppressing sensitivity to pitch motion. A compact, vacuum-compatible verticality monitor based on a silicone-oil level reference and imaging with a CMOS camera is developed to measure and compensate for beam tilt. A new TIA system has been built and is currently being tested. A novel coil motion measurement system is proposed~\cite{liucpem2026}: Three optical sensors are used to measure the displacement of three corners of a triangle connected to the coil, determining $\theta_x$ and $\theta_y$, similar to those systems using three interferometers, and there are two additional channels to measure $x$ and $y$. The optical sensors have a measurement resolution better than 0.3\,$\upmu$m over a 3\,mm range, supporting coil-motion measurement at the $\upmu$m and $\upmu$rad scales. Experimental tests show such a system has good robustness, as it almost never loses signals.

\IEEEpubidadjcol

\section{Conclusion}
This summary presents an overview of the current status of the Tsinghua tabletop Kibble balance, with emphasis on recent progress in its magnetic, mechanical, electrical, and optical subsystems. Future work will focus on integrated weighing and velocity measurements under vacuum, comprehensive uncertainty evaluation, and long-term stability studies. 

\section*{Acknowledgment}
This work was supported by the National Natural Science Foundation of China under Grant 52377011 and in part by the National Key Research and Development Program of China under Grant 2022YFF0708600.


\end{document}